\documentclass[pra,aps,amssymb,twocolumn,showpacs,superscriptaddress]{revtex4}
\usepackage{graphicx} 
\usepackage{amsmath,bm,bbm}
\usepackage{times}
\begin{document}

\newcommand{\proofend}{\hfill\fbox\\\smallskip }
\newcommand{\specnorm}[1]{\lvert\lvert#1\rvert\rvert_\infty}

\title{ Driving non-Gaussian to Gaussian states with linear optics}

\author{Daniel~E.~Browne}
\email{d.browne@ic.ac.uk}
\affiliation{QOLS, Blackett Laboratory, Imperial College London,
Prince Consort Road, London SW7 2BW, UK}

\author{Jens~Eisert}
\affiliation{QOLS, Blackett Laboratory, Imperial College London,
Prince Consort Road, London SW7 2BW, UK}
\affiliation{Institut f{\"u}r Physik, Universit{\"a}t Potsdam,
Am Neuen Palais 10, D-14469 Potsdam, Germany}

\author{
Stefan~Scheel}
\author{ Martin~B.~Plenio}
\affiliation{QOLS, Blackett Laboratory, Imperial College London,
Prince Consort Road, London SW7 2BW, UK}

\date{\today}

\begin{abstract}
We introduce a protocol that maps finite-dimensional pure input states
onto approximately Gaussian states in an iterative procedure. This
protocol can be used to distill highly entangled bi-partite Gaussian
states from a supply of weakly entangled pure Gaussian states. The
entire procedure requires only the use of passive optical elements and
photon detectors that solely distinguish between the presence and
absence of photons.
\end{abstract}


\maketitle

%%%%%%%%%%%%%%%%%%%%%%%%%%%%%%%%%%%%%%%%%%%%%%%%%%%%%%%%%%%%%%%%%%%%%%
\section{Introduction}

Gaussian entangled states   may be   prepared quite simply
in optical systems: one only has to mix a pure squeezed state with
a vacuum state at a beam splitter, both of which are special
instances of Gaussian states in systems with canonical coordinates
\cite{generation,Gauss}. The beam splitter acts as a Gaussian
unitary operation which modifies the quantum state, but does not
alter the Gaussian character of the state. The resulting pure
state is a {\it two-mode squeezed state}. This state may   be
used as the resource   for protocols in quantum information
processing. In fact, teleportation \cite{teleportation}, dense
coding \cite{densecoding} and cryptographic schemes
\cite{cryptography} on the basis of such two-mode squeezed states
have been either studied theoretically or already experimentally
realised.   For the theory of quantum information processing in
systems with canonical degrees Gaussian states play a role closely
analogous to that of entangled states of qubits,   for which
most of the theory of quantum information processing has been
developed.

However, there are significant limits to what accuracy highly
entangled two-mode squeezed states may be prepared and distributed
over large distances. Firstly, the degree of   single-mode
squeezing that   can be achieved limits the degree of two-mode
squeezing of the resulting state. Secondly, decoherence is
unavoidable in the transmission of states through fibres, and the
original highly entangled state will deteriorate into a very
weakly entangled mixed Gaussian state \cite{Stefan}.
For finite-dimensional systems, it has been one of the key
observations   that in fact, from weakly entangled states one
can obtain highly entangled states by means of local quantum
operations supported by classical communication \cite{Bennett} at
the price of starting from a large number of weakly entangled
systems but ending with a smaller number of more strongly
entangled systems.   The term entanglement distillation has
been coined for such procedures. Importantly, such methods
function also as the basis for security proofs of quantum
cryptographic schemes \cite{HansHans}.

  It was generally expected that an analogous procedure should
exists for the distillation of Gaussian states by means of local
Gaussian operations and classical communication only. Surprisingly
however, it was recently proven that this is not the case
\cite{Dist,Dist2}.   For example, no matter how the local
Gaussian quantum operations are chosen, one cannot map a large
number of weakly entangled two-mode squeezed states onto a single
highly entangled Gaussian state. Gaussian quantum operations
\cite{Dist,Dist2,operations} correspond in optical systems to the
application of optical elements such as beam splitters, phase
shifts and $\chi^{(2)}$-squeezers, together with homodyne
detection. All these operations are, to some degree of accuracy,
experimentally accessible. With non-Gaussian quantum operations,
in turn, one can distill finite-dimensional states out of a supply
of Gaussian states \cite{Duan}, but the resulting states are not
Gaussian, and the experimental implementation of the known
protocols constitutes a significant challenge.

\begin{figure}[th]
\centerline{
%        \epsfxsize=7.cm
%       \epsfbox{Figure.eps}
\includegraphics[width=7.5cm]{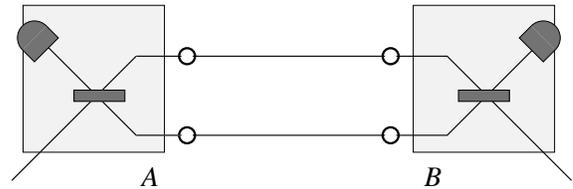}
       }

\vspace{.2cm}
\caption{\label{fig:scheme1} A single step of the protocol. Two pairs
of entangled two-mode states are mixed locally at 50:50 beam splitters
and absence or presence of photons is detected in one of the output
arms on both sides.}
\end{figure}

One may be tempted to think that this observation renders all
attempts to increase the degree of entanglement in Gaussian states
impossible. In this article, however, we discuss the possibility
of obtaining a Gaussian state with arbitrarily high fidelity from
a supply of non-Gaussian states   employing only Gaussian
operations, namely linear optical elements and projections onto
the vacuum.   We describe a protocol that prepares approximate
Gaussian states from a supply of non-Gaussian states,
which shall be called 'Gaussification' from now on. 
The non-Gaussian states
that we use could in particular be obtained from weakly two-mode
squeezed vacua, by the application of a beam splitter and a photon
detector. Together with this step, the proposed procedure offers a
complete distillation procedure of Gaussian states to (almost
exact) Gaussian states, but via non-Gaussian territory.
It is important to note that the protocol introduced
below is by no means restricted to a bi-partite setting. The
bi-partite case is only practically the most important one, as it
allows in effect for distillation of Gaussian states with non-Gaussian
operations. But this method can, in particular,
also be used in a mono-partite setting to approximately obtain a
Gaussian state from a supply of unknown non-Gaussian states.

The paper is organised as follows: First, we will describe the protocol, that generates Gaussian states from a supply of non-Gaussian states. This protocol requires only
passive optical elements and photon detectors that can distinguish
between the absence or presence of photons but that
do not determine their exact number. We then proceed
by discussing the effect of the protocol in more detail.
We will discuss the special case of pure states in Schmidt form as
well as general pure states. The fixed points of the iteration map
will be identified as pure Gaussian states and a proof of convergence
will be given. Finally, we will discuss the feasible preparation of
finite-dimensional states from a supply of pure Gaussian states.

%%%%%%%%%%%%%%%%%%%%%%%%%%%%%%%%%%%%%%%%%%%%%%%%%%%%%%%%%%%%%%%%%%%%%%
\section{The protocol}
\label{sec:protocol}

The protocol is very simple indeed. We start with a supply of
identically prepared bi-partite non-Gaussian states. The overall
protocol then amounts to an iteration of the following basic steps:
\begin{enumerate}
\item  The states will be mixed pairwise locally at
50:50 beam splitters (see Fig.~\ref{fig:scheme1}).
\item On one of the outputs of each beam splitter a photon detector
distinguishes between the absence and presence of photons.
It should be noted that we do not require photon counters that can
discriminate between different photon numbers.
\item In case of absence of photons at  both detectors for a
particular pair one keeps the remaining modes as an input for the next
iteration, otherwise the state is discarded.
\end{enumerate}

This is one iteration of the protocol which we will continue until
we finally end up with a small number of states that closely
resemble Gaussian states. This is clearly a probabilistic
protocol. However, the success probability, as we will see later,
can be quite high.   It should also be noted that the
operations in a successful run are indeed Gaussian operations,
namely the use of linear optical elements and vacuum
projections.  Each of these steps can be realised with
present-day technology.

%%%%%%%%%%%%%%%%%%%%%%%%%%%%%%%%%%%%%%%%%%%%%%%%%%%%%%%%%%%%%%%%%%%%%%
\section{Examples of the protocol}

%%%%%%%%%%%%%%%%%%%%%%%%%%%%%%%%%%%%%%%%%%%%%%%%%%%%%%%%%%%%%%%%%%%%%%
\subsection{Pure states in Schmidt form}

In order to demonstrate the general mechanism, we start by
discussing a particularly simple case,   namely   pure
states in Schmidt form. We do not require any prior knowledge of
the actual un-normalised state vectors except that they can be
expressed in the following form
\begin{eqnarray}
|\psi^{(0)}\rangle = \sum_{n=0}^{\infty} \alpha_{n,n}^{(0)} |n,n\rangle,
\end{eqnarray}
where $\{\alpha_{n,n}^{(0)}\}_{n=0}^{\infty}$ with
$\alpha_{n,n}^{(0)}\!\geq\!0$ are proportional to the real Schmidt
coefficients of the state vector, and
$\{|n\rangle:n\in{\mathbb{N}}\}$ denotes the Fock basis. We only
assume $\alpha_{0,0}^{(0)}\!>\!0$ and it is then convenient to
consider un-normalized states for which we set
$\alpha_{0,0}^{(0)}\!=\!1$. The un-normalized states arising in
later steps $i\!=\!1,2,\ldots$ are characterised by coefficients
$\{\alpha_{n,n}^{(i)}\}_{n=0}^{\infty}$. These coefficients then
become identical to the Schmidt coefficients only after
appropriate normalisation. Starting from two   identical copies
of state vectors that have been obtained in the $i$th step of the
protocol   , i.e.
\begin{eqnarray}
    |\psi^{(i)}\rangle|\psi^{(i)}\rangle
\end{eqnarray}
one obtains after application of the 50:50 beam splitters the
state vector $(\hat U_{12}\otimes \hat
U_{12})|\psi^{(i)}\rangle|\psi^{(i)}\rangle$. Here, the beam
splitter is described by (see, e.g., \cite{vogelwelsch})
\begin{equation}
    \hat{U}_{12} = T^{\hat{n}_1} e^{-R^\ast \hat{a}_2^\dagger \hat{a}_1}
    e^{R \hat{a}_2 \hat{a}_1^\dagger} T^{-\hat{n}_2} \,,
\end{equation}
where $\hat{U}_{12}$ acts on the amplitude operators of the field
modes as
\begin{equation}
    \hat{U}_{12} \left(
    \begin{array}{c}
    \hat{a}_1 \\ \hat{a}_2
    \end{array} \right) \hat{U}_{12}^\dagger =
    \left( \begin{array}{cc}
    T & R \\ -R^\ast & T^\ast
    \end{array} \right)
    \left(
    \begin{array}{c}
    \hat{a}_1 \\ \hat{a}_2
    \end{array} \right)
\end{equation}
where we set $T\!=\!R\!=\!1/\sqrt{2}$. The resulting un-normalised state
vector, conditional on vacuum outcomes in both detectors, is given by
\begin{eqnarray}
    |\psi^{(i+1)}\rangle &:= &\langle 0,0 | (\hat U_{12}\otimes \hat
    U_{12})|\psi\rangle|\psi\rangle\nonumber\\
    &=&\sum_{n=0}^{\infty} \left[2^{-n}\sum_{r=0}^{n} {n \choose r}
    \alpha_{r,r}^{(i)} \alpha_{n-r,n-r}^{(i)} \right]
    |n,n\rangle\nonumber\\
    &=& \sum_{n=0}^{\infty}  \alpha_{n,n}^{(i+1)} |n,n\rangle,
\end{eqnarray}
where
\begin{eqnarray}
\label{eq:recursion}
    \alpha_{n,n}^{(i+1)}:= 2^{-n}\sum_{r=0}^{n} {n \choose r}
    \alpha_{r,r}^{(i)} \alpha_{n-r,n-r}^{(i)}
\end{eqnarray}
for $n=0,1,\ldots$.%%%%%%%%%%%%%%%
The probability of vacuum outcomes being detected in
both modes is $\langle\psi^{(i+1)}|\psi^{(i+1)}\rangle/\lvert\langle\psi^{(i)}|\psi^{(i)}\rangle\rvert^2$. %%%%%%
The protocol is a Gaussian quantum operation, in the sense that it
is a completely positive map that maps all Gaussian states onto
Gaussian states. The interesting feature is that by repeated
application it also maps non-Gaussian states arbitrarily close to
Gaussian states, as will be demonstrated below.

In effect, in each iteration one maps one sequence of coefficients
$\alpha^{(i)}\!=\!\{\alpha_{n,n}^{(i)}\}_{n=0}^{\infty}$ onto another
sequence $\alpha^{(i+1)}\!=\!\{\alpha_{n,n}^{(i+1)}\}_{n=0}^{\infty}$,
defining the map $\Phi$ via
\begin{eqnarray}
    \alpha^{(i+1)}=: \Phi(\alpha^{(i)}) \,.
\end{eqnarray}
In the following we use the notation
$\Phi^{(1)}\!=\!\Phi$ and
$\Phi^{(i+1)}\!=\!\Phi\circ\Phi^{(i)}$ for $i=0,1,...$.
The main observation is that in fact, provided
$\alpha_{1,1}^{(0)}\!<\!\alpha_{0,0}^{(0)}$, the sequence
of coefficients $\{\alpha^{(i)}\}_{i=1}^{\infty}$ converges to a
distribution corresponding to a Gaussian state, in this special
case a two-mode squeezed vacuum.

In other words, although the initial state was not Gaussian, but
say, a state corresponding to a finite-dimensional state vector of
the form
\begin{equation}
    |\psi^{(0)}\rangle = |0,0\rangle + \alpha_{1,1}^{(0)} |1,1\rangle,
\end{equation}
where $\alpha_{1,1}^{(0)}\in [0,1)$, after a number of steps the
resulting state is Gaussian to a high degree of accuracy. We will
first show that this convergence is a general feature of this
protocol, and we will then discuss the consequences. We start by
demonstrating that those distributions associated with pure
Gaussian states are fixed points of the map $\Phi$.

\vspace*{1ex}
\noindent
{\bf Proposition 1. -- } {\it The distributions
$\alpha=\{\alpha_{n,n}\}_{n=0}^{\infty}$ of the form
\begin{eqnarray}
    \alpha_{n,n}= \lambda^{n},% \,\,\,\lambda\in[0,1)
\end{eqnarray}
$\lambda\geq 0$,   corresponding to two-mode squeezed 
states, are the only fixed points of the map $\Phi$.}

\vspace*{1ex}
{\it Proof.} This can be immediately derived from the definition of
$\Phi$: Let us assume that
\begin{equation}\label{prop1proof}
        \alpha\!=\!\Phi(\alpha)
\end{equation}
holds.
It can be verified by substitution that $\alpha_{n,n}=\lambda^n$ is a solution of this equation. The uniqueness of this solution can be verified by observing that  Eq.~\eqref{prop1proof} also implies  $\alpha_{0,0}\!=\!\alpha_{0,0}^{2}$, that is,
$\alpha_{0,0}\!=\!1$. Then $\alpha_{1,1}$ is a free parameter and once set (i.e. as $\alpha_{1,1}=\lambda$) the remaining coefficients are uniquely determined.\proofend

These coefficients, for $\lambda\in[0,1)$, in turn correspond exactly
to two-mode pure Gaussian states.
%, if $\sum_{n=0}^\infty \alpha_{n,n}^2<\infty$.
If $\lambda$ lies outside this
range, the state is not normalizable.
% which is a special case of
%Proposition~3 below.
The next Proposition states that those distributions
associated with Gaussian states are not only fixed points of the
map $\Phi$, but provided $\alpha_{0,0}^{(0)}\!\neq\!0$, each sequence of
coefficients converges to such a fixed point.

\vspace*{1ex}
\noindent
{\bf Proposition 2. -- } {\it Let
$\alpha^{(0)}\!=\!\{\alpha_{n,n}^{(0)}\}_{n=0}^{\infty}$ with
$a_{0,0}^{(0)}=1$ and
$0\!\leq \!\alpha_{1,1}^{(0)}\!<\!1$. Then
\begin{eqnarray}
    \lim_{i\rightarrow \infty} \alpha^{(i)}_{n,n} =
    \alpha^{(\infty)}_{n,n}
\end{eqnarray}
for all $n=0,1,...$,
where $\alpha^{(\infty)}$ is a distribution of the type of
Proposition 1.}

\vspace*{1ex}
{\it Proof.} As before, let us set
$\alpha^{(i)}\!:=\!\Phi^{(i)}(\alpha^{(0)})$
for $i\!=\!1,2,\ldots$.
The first step is to see that
\begin{eqnarray}
    \frac{\alpha_{1,1}^{(i+1)}}{\alpha_{0,0}^{(i+1)}} =
    \frac{\alpha_{1,1}^{(i)}}{\alpha_{0,0}^{(i)}}= \alpha_{1,1}^{(0)}
\end{eqnarray}
for all $i\!=\!0,1,\ldots$. Let us first assume that $\alpha_{1,1}^{(0)}>0$.
Then, as can be seen from the definition of $\Phi$,
\begin{eqnarray}
    \alpha_{2,2}^{(i+1)}\alpha_{1,1}^{(i)}  = \frac{1}{2}\left(
     \alpha_{2,2}^{(i) }+
    \alpha_{1,1}^{(0)}\alpha_{1,1}^{(i)}\right)\alpha_{1,1}^{(i+1)}.
\end{eqnarray}
Hence, as $\alpha_{1,1}^{(i)}=\alpha_{1,1}^{(0)}>0$ for all $i=0,1,...$,
\begin{eqnarray}
    \lim_{i\rightarrow \infty} \frac{\alpha_{2,2}^{(i)}}{\alpha_{1,1}^{(i)}}=
    \alpha_{1,1}^{(0)}.
\end{eqnarray}
Now let us assume that already $\alpha_{n-1,n-1}^{(i)}>0$ for all $i=0,1,...$ and
\begin{eqnarray}
    \lim_{i\rightarrow \infty}
    \frac{\alpha_{n,n}^{(i)}}{\alpha_{n-1,n-1}^{(i)}}=
    \alpha_{1,1}^{(0)}
\end{eqnarray}
for some $n\!=\!1,2,\ldots$. Then, from
\begin{eqnarray}
    \frac{\alpha_{n+1,n+1}^{(i+1)}}{\alpha_{n,n}^{(i+1)}}&=& \frac{1}{2}
    \frac{\sum_{r=0}^{n+1} \alpha_{r,r}^{(i)} \alpha_{n-r+1,n-r+1}^{(i)}
    {n+1 \choose r}}{\sum_{r=0}^{n}
    \alpha_{r,r}^{(i)} \alpha_{n-r,n-r}^{(i)} {n \choose r}}
\end{eqnarray}
it follows after a few steps that $a_{n,n}^{(i)}>0$ for all $i=0,1,...$, and
\begin{equation}
\begin{split}
    &\lim_{i\rightarrow \infty}
    \frac{\alpha_{n+1,n+1}^{(i+1)}}{\alpha_{n,n}^{(i+1)}}=\\
    &\lim_{i\rightarrow \infty} \frac{1}{2^{n+1}}\left[
    \frac{2\alpha_{n+1,n+1}^{(i)}}{\alpha_{n,n}^{(i)}} +
    (2^{n+1}-2) \alpha_{1,1}^{(0)}
    \right],
\end{split}
\end{equation}
which means that
\begin{eqnarray}
    \lim_{i\rightarrow \infty}
    \frac{\alpha_{n+1,n+1}^{(i+1)}}{\alpha_{n,n}^{(i+1)}}=\alpha_{1,1}^{(0)}.
\end{eqnarray}
Hence, by induction we find that the ratios of $\alpha_{n+1,n+1}^{(i)}$ and
$\alpha_{n,n}^{(i)}$ converge to the ratio of $0\!<\!\alpha_{1,1}^{(0)}\!<\!1$ and
$\alpha_{0,0}^{(0)}\!=\!1$ as $i\!\to\!\infty$. This means that the
coefficients correspond to a Gaussian state as specified in
Proposition 1. In case that $\alpha_{1,1}^{(0)}=0$ an analogous argument can be
applied in order to arrive at $\alpha_{0,0}^{(i)}=1$ for all $i=0,1,...$ and
\begin{equation}
         \lim_{i\rightarrow \infty}\alpha_{n,n}^{(i)}=0
\end{equation}
for all $n\!=\!1,2,\ldots$.
\proofend

\begin{figure}[ht]
\centerline{\includegraphics[width=7cm]{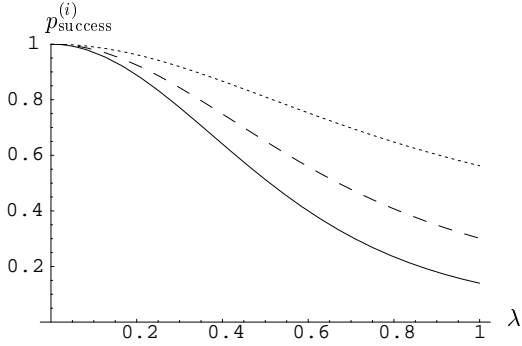}}
\caption{\label{fig:prob3} Success probability $p_{\text{success}}^{(i)}$
after $i\!=\!1$ (dotted line),
$i\!=\!2$ (dashed line) and $i\!=\!3$ (solid line) iteration steps,
where the initial states were
$\propto|0,0\rangle+\lambda|1,1\rangle$.}
\end{figure}

\begin{figure}[ht]
\centerline{\includegraphics[width=7cm]{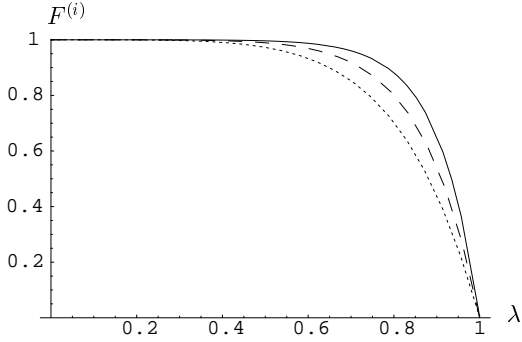}}
\caption{\label{fig:fid3} Fidelity $F^{(i)}$ of the approximately
Gaussian state after $i\!=\!1$ (dotted line), $i\!=\!2$ (dashed line) and
$i\!=\!3$ (solid line) iterations where the initial states were
$\propto|0,0\rangle+\lambda|1,1\rangle$.}
\end{figure}

This shows formally that the (pointwise)
convergence to an effectively Gaussian
state is generic \cite{Gen}. 
Putting aside
the restriction that $\alpha_{0,0}^{(0)}=1$,
three cases shall be discussed in more detail.
\begin{enumerate}
    \item If $\alpha_{0,0}^{(0)}\!>\!0$ and
    $\alpha_{1,1}^{(0)}\!<\!\alpha_{0,0}^{(0)}$,
    then the states converge to a Gaussian state.
    \item A special instance is when $\alpha_{0,0}^{(0)}\!>\!0$, but
    $\alpha_{1,1}^{(0)}\!=\!0$. Then the states converge to a Gaussian state,
    but to the product of two vacua.
    \item If $\alpha_{0,0}^{(0)}\leq \alpha_{1,1}^{(0)}$, then the
    sequence does
    not converge to a sequence of coefficients corresponding to
    a Gaussian state. In particular, this is always the case when
    \begin{equation}
        \alpha_{0,0}^{(0)}\!=\!0.
    \end{equation}
    This follows immediately from
    Eq.~(\ref{eq:recursion}) as $\alpha_{0,0}^{(i)}\!=\!0$ for all $i$.
\end{enumerate}
In practice,  one can actually expect a state that is very
close to a Gaussian state already after a very small number of
steps, say, three or four steps. As has already been mentioned,
the whole scheme is probabilistic. That is, the success
probability of actually obtaining the desired state is always less
than one. In Fig.~\ref{fig:prob3} we show the total  probability of
success, $p_{\text{success}}^{(i)}$, and in Fig.~\ref{fig:fid3} the
corresponding fidelity $F^{(i)}$, i.e. the overlap with the Gaussian
state to which the protocol converges, after $i\!=\!1,2,3$ iteration
steps. Here, we started with coefficients $\alpha_{0,0}^{(0)}\!=\!1$ and
$\alpha_{1,1}^{(0)}\!=\!\lambda$.
We see that for a large range of values for $\lambda$ the fidelity
is just below unity, and for $\lambda\!=\!0.5$ the probability of
success is still above 0.5.

\subsection{General pure states}

Suppose now we have a supply of pure states with state vectors
of the general form
\begin{equation}
|\psi^{(0)}\rangle=\sum_{m,n=0}^\infty
\alpha^{(0)}_{m,n} |{m,n}\rangle,
\end{equation}
where $ \alpha^{(0)}_{m,n}\in{\mathbbm{C}}$ for all $n,m$.
If the  procedure described in Sec.~\ref{sec:protocol} is carried out, using
50:50 beam splitters with appropriate phases, such that
$T\!=\!R\!=\!1/\sqrt{2}$, then, for a large class of input states, after
 repeated
iterations of the protocol, a state closely approximating a Gaussian state will be obtained.
If the identical retained states after $i$ iterations of the procedure
are labeled
\begin{equation}
|\psi^{(i)}\rangle\!=\!\sum_{m,n} \alpha^{(i)}_{m,n} |m,n\rangle,
\end{equation}
we can describe each iteration in terms of the
following recurrence relation,
\begin{equation}
\label{recrels}
\begin{split}
\alpha_{m,n}^{(i)}
\longmapsto
%\stackrel{\Phi}{\longmapsto}
\alpha_{m,n}^{(i+1)}&=2^{-\frac{m+n}{2}}\sum_{r=0}^m\sum_{s=0}^n
(-1)^{(m+n)-(r+s)}\\
&\times 
\alpha_{r,s}^{(i)}\alpha_{m-r,n-s}^{(i)}
\left[{{m \choose r}{n \choose s}}\right]^{1/2},
\end{split}
\end{equation}
where again
\begin{equation}
\alpha^{(i+1)}= \Phi(\alpha^{(i)}),
\end{equation}
with $\alpha^{(i)}=\{\alpha^{(i)}_{n,m}\}_{n,m=0}^\infty$ for $i=0,1,...$.
We will in the following write
\begin{equation}
        \alpha_{n,m}^{(\infty)}:= \lim_{i\rightarrow\infty} \alpha_{n,m}^{(i)},
\end{equation}
whenever this limit exists.
The fixed points of $\Phi$, characterised by
$\alpha_{m,n}^{(\infty)}\in {\mathbbm{C}}$,
correspond to states
which are unchanged by one or more iterations of the procedure, and
satisfy $\Phi(\alpha^{(\infty)})\!=\!\alpha^{(\infty)}$, thus
\begin{equation}
\label{fixedeq}
\begin{split}
\alpha_{m,n}^{(\infty)}&=2^{-\frac{m+n}{2}}\sum_{r=0}^m\sum_{s=0}^n
(-1)^{(m+n)-(r+s)}\\&
\times \alpha_{r,s}^{(\infty)}\alpha_{m-r,n-s}^{(\infty)}
\left[{{m \choose r}{n \choose s}} \right]^{1/2}
\end{split}
\end{equation}
for all $n,m$.
We immediately see that
\begin{equation}
\alpha_{0,0}^{(\infty)}\!=\!(\alpha_{0,0}^{(\infty)})^2
\end{equation}
and thus $\alpha_{0,0}^{(\infty)}\!=\!1$. (The other possibility,
$\alpha_{0,0}^{(\infty)}\!=\!0$ leads to the trivial solution
$\alpha_{m,n}^{(\infty)}\!=\!0$ for all $m,n$.)
We also find that the coefficients $\alpha_{1,1}^{(\infty)}$,
$\alpha_{2,0}^{(\infty)}$ and  $\alpha_{0,2}^{(\infty)}$ are the only
free parameters. When these values are specified, all other
coefficients are determined.
The general solution of Eq.~\eqref{fixedeq} is
%\begin{equation}
%    \alpha_{2m,2n+1}^{(\infty)}=\alpha_{2m+1,2n}^{(\infty)}=0 \,,
%\end{equation}
\begin{equation}
    \alpha_{2m,2n+1}^{(\infty)}=\alpha_{2m+1,2n}^{(\infty)}=0\,,
\end{equation}
\begin{eqnarray}
\lefteqn{
    \alpha_{2m,2n}^{(\infty)}=\sqrt{(2m)!}\sqrt{(2n)!} }
\nonumber \\ &&
    \times \sum_{0\leq s\leq m;s\leq n}
    \left[\frac{\gamma_{12}^{2s}}{(2s)!}
    \frac{\left(\gamma_{1}/{2}\right)^{m-s}}{(m-s)!}
    \frac{\left(\gamma_{2}/{2}\right)^{n-s}}{(n-s)!}\right] \,,
\end{eqnarray}
\begin{eqnarray}
\lefteqn{
    \alpha_{2m+1,2n+1}^{(\infty)}=\sqrt{(2m+1)!}\sqrt{(2n+1)!} }
\nonumber \\ &&
    \times \sum_{0\leq s\leq m;s\leq n}
    \left[\frac{\gamma_{12}^{2s+1}}{(2s+1)!}
    \frac{\left(\gamma_{1}/{2}\right)^{m-s}}{(m-s)!}
    \frac{\left(\gamma_{2}/{2}\right)^{n-s}}{(n-s)!}\right] \,,
\end{eqnarray}
where the coefficients $\gamma_1,\gamma_2$ and $\gamma_{12}$ are
usefully expressed as elements of the symmetric $2\times2$ matrix
\begin{equation}
\bm{\Gamma}=\left(\begin{array}{cc}
\gamma_1&\gamma_{12}\\
\gamma_{12}&\gamma_2
\end{array}\right)
\label{gamma}
\end{equation}
and are determined uniquely by the
free parameters $\alpha_{2,0}^{(\infty)},\alpha_{0,2}^{(\infty)}$
and $\alpha_{1,1}^{(\infty)}$. A specific form for this correspondence is
given in Proposition 4. The coefficients $\alpha_{mn}^{(\infty)}$
determine an un-normalized state vector
$|\psi(\bm{\Gamma})\rangle$. In
the Fock state representation this state vector
is given by
\begin{equation}
    \label{fixedpointstates}
    |\psi(\bm{\Gamma})\rangle=\hat{Q}(\bm{\Gamma})|0,0\rangle,
\end{equation}
where the operator $\hat{Q}(\bm{\Gamma})$ is expressed in terms of
$\bm{\Gamma}$ and the vector
$\hat{\bm{a}}^\dag\!=\!(\hat{a}^\dag_1,\hat{a}^\dag_2)^T$ as
\begin{equation}
    \hat{Q}(\bm{\Gamma})=\exp\left[\frac{1}{2}(\hat{\bm{a}}^\dag)^T
    \bm{\Gamma} (\hat{\bm{a}}^\dag)\right] \,.
\end{equation}
The state vectors
$|\psi(\bm{\Gamma})\rangle$ are not normalized, and the
requirement that they be normalizable, i.e.
$\langle\psi(\bm{\Gamma})|\psi(\bm{\Gamma})\rangle$ is finite,
places a restriction on $\bm{\Gamma}$. The following
proposition takes its most concise form when we use the spectral
norm which is defined as \cite{Horn}
\begin{equation}
\specnorm{\bm{X}}=\sqrt{\lambda_{\text{max}}}
\end{equation}
where $\lambda_{\text{max}}$ is the largest eigenvalue of $\bm{X}
\bm{X}^\dag$.\\

\vspace*{1ex}
\noindent {\bf Proposition 3. -- } {\it If and only if
$\specnorm{\bm{\Gamma}}\!<\!1$, then
$|\psi(\bm{\Gamma})\rangle\!:=\!\hat{Q}(\bm{\Gamma})|0,0\rangle$ is
normalizable and represents a pure Gaussian state.}

\smallskip

{\it Proof:} The matrix $\bm{\Gamma}$ in Eq.~(\ref{gamma}) is a
complex symmetric $2\times 2$-matrix. Following Takagi's Lemma
\cite{Horn}, there exists a unitary matrix $\bm{U}$ such that
\begin{equation}
\bm{U}^T \bm{\Gamma} \bm{U}\!=:\!\bm{\Delta},
\end{equation}
where $\bm{\Delta}$ is a
diagonal matrix the entries of which are the eigenvalues of
$\sqrt{\bm{\Gamma}\bm{\Gamma}^{\dagger}}$.
With $\hat{\bm{b}}\!:=\!\bm{U}\hat{\bm{a}}$ we have
\begin{equation}
    |\psi(\bm{\Gamma})\rangle =
    \exp\left[\frac{1}{2}(\hat{\bm{b}}^\dag)^T
    \bm{\Delta} (\hat{\bm{b}}^\dag)\right] |0,0\rangle \, .
\end{equation}
Because the $\hat{b}_1$ and $\hat{b}_2$ commute, this is a tensor
product of two single-mode Gaussian states. It is now straightforward
to show that the single mode state vectors are normalizable if and only if both
diagonal elements of
$\bm{\Delta}$ are smaller than one. Then, each of the
modes is in a single-mode squeezed state \cite{Ma90}. The
transformation $\hat{\bm{a}}\!\longmapsto\!\bm{U}\hat{\bm{a}}$ represents
a beam-splitter transformation mapping the original modes $\hat{\bm{a}}$
onto the modes $\hat{\bm{b}}$, i.e. it is a passive transformation.
Hence, the resulting state vector Eq.~(\ref{fixedpointstates}) is also
normalizable. \proofend

In fact, as can be shown, the state vector $\hat{Q}(\bm{\Gamma})|0,0\rangle$ is, apart
from normalisation, equal to the state vector of the
two-mode squeezed vacuum state
$\hat{S}(\bm{Z})|0,0\rangle$, where
\begin{equation}
\hat{S}(\bm{Z})=\exp\left[\frac{1}{2}(\hat{\bm{a}}^\dag)^T \bm{Z}
(\hat{\bm{a}}^\dag)-\frac{1}{2}(\hat{\bm{a}})^T \bm{Z}^\dag (\hat{\bm{a}})\right] \,.
\end{equation}
$\hat{S}(\bm{Z})$ is a generalized two-mode squeezing operator \cite{Ma90},
\begin{equation}
\bm{Z}=-\left( \begin{array}{cc}
\zeta_1 & \zeta_{12}\\ \zeta_{12} & \zeta_{2}
\end{array} \right),
\end{equation}
where $\bm{Z}\!=\!\mbox{\rm arctanh}(\bm{r_\Gamma})e^{i \bm{\theta_\Gamma}}$
with the polar decomposition
\begin{equation}
\bm{\Gamma}\!=\!\bm{r_\Gamma}e^{i \bm{\theta_\Gamma}}.
\end{equation}

\vspace*{1ex}
\noindent
{\bf Proposition 4. --} {\it Suppose we are given a supply of
identical two-mode pure states with state vectors
$|\psi^{(0)}\rangle\!=\!\sum_{m,n}\alpha_{m,n}^{(0)}|m,n\rangle$, and let
\begin{equation}
\label{gammavalues}
\bm{\Gamma}:=\left(\begin{array}{cc}
\sqrt{2}\beta_{2,0}-\beta_{1,0}^2 &
\beta_{1,1}-\beta_{1,0} \beta_{0,1}\\
\beta_{1,1}-\beta_{1,0} \beta_{0,1} &
\sqrt{2} \beta_{0,2}-\beta_{0,1}^2
\end{array} \right),
\end{equation}
where $\beta_{m,n}\!:=\!\alpha_{m,n}^{(0)}/\alpha_{0,0}^{(0)}$.
If $\specnorm{\bm{\Gamma}}\!<\!1$ then 
\begin{equation}
        \lim_{i\rightarrow \infty} \alpha_{m,n}^{(i)} = 
        \alpha_{m,n}
\end{equation}
for all $n,m=0,1,...$, where 
\begin{equation}
        \alpha_{m,n} := \langle m,n| \hat{Q}(\bm{\Gamma})|0,0\rangle.
\end{equation}
}

{\it Proof.} To make the proof simpler, we shall use
$\alpha_{0,0}^{(0)}\!=\!1$ as above. This is merely a change of
normalization and does not alter the general validity of the
argument. Before proving the convergence of all coefficients
$\alpha_{m,n}^{(i)}$ under $\Phi$ to the fixed point
$\alpha_{m,n}^{(\infty)}$ as $i\!\to\!\infty$, let us first show that a
certain subset of coefficients actually reach their final value after
a single iteration of $\Phi$.

The coefficients $\alpha_{2m+1,2n}^{(1)}$ and
$\alpha_{2m,2n+1}^{(1)}$ reach zero, their fixed point, after a
single iteration corresponding to $i=1$,
for all $m,n$. To see this, note that
in the following equation,
\begin{equation}
\label{fixedeq1}
\begin{split}
\alpha_{m,n}^{(1)}&= 2^{-\frac{m+n}{2}}\sum_{r=0}^m\sum_{s=0}^n
(-1)^{(m+n)-(r+s)}\\& \times
\alpha_{r,s}^{(0)}\alpha_{m-r,n-s}^{(0)} \left[{{m \choose r}{n
\choose s}}\right]^{1/2}
\,,
\end{split}
\end{equation}
renaming the summation indices $(r,s)\!\mapsto\!(m-r,n-s)$,
yields an identical sum except for an overall factor of
$(-1)^{m+n}$. Consequently, for odd values of $m\!+\!n$ the whole sum
must vanish and coefficients of the form $\alpha^{(1)}_{2m+1,2n}$ and
$\alpha^{(1)}_{2m,2n+1}$ vanish after a single iteration step.
As a consequence of this, the coefficients $\alpha_{1,1}^{(i)}$,
$\alpha_{2,0}^{(i)}$ and $\alpha_{0,2}^{(i)}$ also do not change after
one iteration. For example,
\begin{equation}
\alpha_{1,1}^{(i+1)}=
\alpha_{1,1}^{(i)}-\alpha_{0,1}^{(i)}\alpha_{1,0}^{(i)}
\end{equation}
for all $i=0,1,...$.
Similarly, $\alpha_{2,0}^{(1)}$ and
$\alpha_{0,2}^{(1)}$ also assume
their respective fixed points after
the first iteration, and thus the matrix $\bm{\Gamma}$ is determined
to be as in Eq.~(\ref{gammavalues}).

Now let us show that all coefficients $\alpha_{m,n}^{(i)}$ do indeed
converge to their respective fixed points $\alpha_{m,n}^{(\infty)}$ as
$i\!\to\!\infty$.
The recurrence relations in Eq.~\eqref{recrels} can be
re-written as
\begin{equation}
\begin{split}
\alpha_{m,n}^{(i+1)}&=2^{-\frac{m+n}{2}}
\biggl[2\alpha_{m,n}^{(i)}+
\hspace*{-3ex}
\underbrace{\sum_{r=0}^m\sum_{s=0}^n}_{(r,s)\neq(0,0)\neq(m,n)}
\hspace*{-3ex}
(-1)^{(m+n)-(r+s)}\\&\qquad \times
\alpha_{r,s}^{(i)}\alpha_{m-r,n-s}^{(i)}
\left[{{m \choose r}{n \choose s}}\right]^{1/2}\biggr] \,.
\end{split}
\end{equation}
%After the first iteration $\alpha_{0,1}^{(i)}$ and $\alpha_{1,0}^{(i)}$
%are zero.
%
Let us assume that all coefficients $\alpha_{r,s}^{(i)}$, where
$r\!\leq\!m$, $s\!\leq\!n$ but $r\!+\!s\!<\!m\!+\!n$, do converge to
the fixed points $\alpha_{r,s}^{(\infty)}$ as $i\to \infty$. Then
\begin{equation}
\label{limrecrel}
\begin{split}
& \lim_{i\to\infty} \alpha_{m,n}^{(i+1)}=
2^{(1-\frac{m+n}{2})}\lim_{i\to\infty} \alpha_{m,n}^{(i)}
\\&+2^{(-\frac{m+n}{2})}
\hspace*{-3ex}
\underbrace{\sum_{r=0}^m\sum_{s=0}^n}_{(r,s)\neq(0,0)\neq(m,n)}
\hspace*{-3ex}
(-1)^{(m+n)-(r+s)}\\& \times
\alpha_{r,s}^{(\infty)}\alpha_{m-r,n-s}^{(\infty)}
\left[{{m \choose r}{n \choose s}}\right]^{1/2} \,.
\end{split}
\end{equation}
Now let us use the substitution
$\delta_{m,n}^{(i)}\!:=\!\alpha_{m,n}^{(i)}\!-\!\alpha_{m,n}^{(\infty)}$ and
we obtain, using Eq.~(\ref{fixedeq}),
\begin{equation}
\lim_{i\to\infty}\delta_{m,n}^{(i+1)}=
2^{(1-\frac{m+n}{2})}\lim_{i\to\infty} \delta_{m,n}^{(i)} \,.
\end{equation}
We see that $\delta_{m,n}^{(i)}$ converges to zero as long as
\begin{equation}
2^{(1-\frac{m+n}{2})}<1,
\end{equation}
which is the case whenever $m+n>2$. However,
since we have already shown that all coefficients $\alpha_{m,n}^{(i)}$,
where $m+n\leq2$, i.e.
$\alpha_{0,0}^{(i)}$, $\alpha_{0,1}^{(i)}$, $\alpha_{1,0}^{(i)}$,
$\alpha_{1,1}^{(i)}$,
$\alpha_{0,2}^{(i)}$ and $\alpha_{2,0}^{(i)}$, converge to a final
value after a single iteration, the convergence of all other
coefficients follows by induction.
Note that whenever $\specnorm{\bm{\Gamma}}\!\ge\!1$, although the
coefficients individually converge to their respective fixed points,
the state as a whole does not, since $\hat{Q}(\bm{\Gamma})|0,0\rangle$
is not a normalizable state vector.\proofend

%%%%%%%%%%%%%%%%%%%%%%%%%%%%%%%%%%%%%%%%%%%%%%%%%%%%%%%%%%%%%%%%%%%%%%
\section{Generation of the initial states from Gaussian states}
\label{Procrustean}

So far we did not specify where the supply of initial states should
come from.
In fact, one could use two (weakly) entangled Gaussian states and
feed them into one of the iteration components shown in
Fig.~\ref{fig:scheme1}. Then, instead of retaining the state in the case of
measuring the vacuum, we now retain the state whenever \textit{any}
nonzero photon number is obtained. Again, only detectors that
distinguish between absence or presence of photons are needed. Let us
start with a supply of two-mode squeezed vacuum states the state
vectors of which can be written in Schmidt basis as
\begin{equation}
\label{eq:tmsv}
|\psi_q \rangle =\sqrt{1-q^2} \sum\limits_{n=0}^\infty q^n
|n,n\rangle \,,
\end{equation}
with $q\in[0,1)$.

  In general, it will be easier to generate two-mode squeezed states with low values of $q$ in
an experiment, and using the following simple protocol one can use a supply of such states to generate a supply of non-Gaussian states which, when used as the input of the procedure described in Sec.~\ref{sec:protocol}, lead to the generation of two-mode squeezed states with much  higher $q$.

Let us feed two copies of the state of the form
as in Eq.~(\ref{eq:tmsv})  with $q\ll 1$ 
into the device schematically depicted in Fig.~\ref{fig:scheme1} and
retain those outcomes that correspond to a `click' in both
detectors. It does not matter how many photons have been
measured, and we do not assume that a different classical signal is
associated with different photon numbers.
The projection operator~\cite{Bartlett02} describing this process is
\begin{equation}
\label{eq:projector}
\hat{P} = (\hat{\openone}-|0\rangle\langle 0|)\otimes
(\hat{\openone}-|0\rangle\langle 0|) \,.
\end{equation}
Although the vacuum projection (as well as the identity operation) are
Gaussian, the difference of them is not, and, indeed, we find that when the states used in the protocol have sufficiently small $q$, then this projection approximates $|1\rangle\langle 1|\otimes|1\rangle\langle 1|$
 with high accuracy. Thus, we are not in the
situation as in Refs.~\cite{Dist,Dist2}.
Acting with (\ref{eq:projector}) on two copies of the state
(\ref{eq:tmsv}) after rotating them at the beam splitters gives the
non-Gaussian state with un-normalised
state vector
\begin{eqnarray}
\lefteqn{
|\Psi(q; T_A,R_A;T_B,R_B)\rangle :=} \nonumber \\ &&
\hat{P}\left[
\hat{U}_{12}(T_A,R_A) \otimes \hat{U}_{12}(T_B,R_B) \right]
| \psi_q \rangle^{\otimes 2}\,,
\end{eqnarray}
where again
\begin{equation}
\hat{U}_{12}(T,R) = T^{\hat{n}_1} e^{-R^\ast \hat{a}_2^\dagger \hat{a}_1}
e^{R \hat{a}_2 \hat{a}_1^\dagger} T^{-\hat{n}_2}
\end{equation}
and $T_A,T_B,R_A,R_B\in{\mathbbm{C}}$ with
\begin{equation}
|T_A|^2 + |R_A|^2=|T_B|^2 + |R_B|^2 =1.
\end{equation}
For simplicity of notation, let
\begin{eqnarray}
    &&\omega(q; T_A,R_A;T_B,R_B):=\\
    &&\frac{\text{tr}_{M}[|\Psi(q; T_A,R_A;T_B,R_B  )\rangle \langle
    \Psi(q; T_A,R_A;T_B,R_B ) |]}
    {\text{tr}[
    |\Psi(q; T_A,R_A;T_B,R_B  )\rangle \langle
    \Psi(q; T_A,R_A;T_B,R_B ) |
    ]}\nonumber
\end{eqnarray}
be the normalised state after application of the beam splitters and
the two projections, where $\text{tr}_{M}$ is the partial trace over the measured modes.
The most appropriate choice for the reflectivities and
transmittivities clearly depends on the value of $q$ and on the figure
of merit of how one quantifies the quality of the output
state. However, when $q\!\in\![0,1)$ is very small, the output state can be
made arbitrarily close to a maximally entangled state 
\begin{equation}
    \rho^+= \frac{1}{\sqrt{2}}\big[
    |0,0\rangle+ e^{-i \phi} |1,1\rangle\big]\big[\langle 0,0|+e^{i\phi}\langle 1,1| \big]
\end{equation}
in $2\times 2$ dimensions. Where the phase $e^{i\phi}$ depends on the phases of $T$ and $R$ in the beam splitter chosen. More precisely,
\begin{equation}
    \lim_{q\rightarrow 0}\left\|\omega(q; t(q),r(q);0,1)
     - \rho^+ \right\|_1=0,
\end{equation}
where
\begin{eqnarray}
    \lvert t(q)\rvert &:= & \left\lvert\frac{1-(1+8 q^2)^{1/2} }{4q }\right\rvert,\\
    \lvert r(q)\rvert &:=& \left[1- \lvert t(q)\rvert^2\right]^{\frac{1}{2}},
\end{eqnarray}
and $\|.\|_1$ denotes the trace-norm \cite{Horn}. In other words,
in the limit of very small two-mode squeezing the maximally
entangled state can be obtained to a high degree of accuracy. So
the appropriate choice for the beam splitters on one side does
depend on the value of $q$, whereas the beam splitter on the other
side becomes redundant. In a similar manner, one can generate
states     of the form $|0,0\rangle + \alpha_{1,1}^{(0)}
|1,1\rangle$. If one does not care about the phase of
$\alpha_{1,1}$, the correct choice for the above transmittivities
and reflectivities is then
\begin{eqnarray}
   \lvert t(q)\rvert &:= &\Biggl| \frac{\lvert\alpha_{1,1}^{(0)}\rvert -\left[\lvert\alpha_{1,1}^{(0)}\rvert^2+8
        q^2\right]^{1/2} }{4q }\Biggr|
\end{eqnarray}
and $\lvert r(q)\rvert\!:=\![1- \lvert  t(q)\rvert^2]^{\frac{1}{2}}$.   This analysis shows that with
the help of passive optical elements and photon detectors, quantum
states of the
appropriate kind can in fact be prepared. There is, however,
a trade-off concerning accuracy of the protocol and success probability:
For any finite $q$,
the resulting states are not exactly pure, whereas the probability of success
(such that the non-vacuum outcome is obtained in both detectors)
is a monotone decreasing function of $q$.

The resulting states of this protocol can then form the starting
point of the generation of Gaussian states via the protocol in Sec.~\ref{sec:protocol}.
In effect, this scheme allows one to generate
approximate Gaussian states (in fact, two-mode squeezed vacua) with
a higher $q$ than the initial supply, which is nothing other than a distillation
procedure.
%%%%%%%%%%%%%%%%%%%%%%%%5

\begin{figure}[ht]
\centerline{\includegraphics[width=7cm]{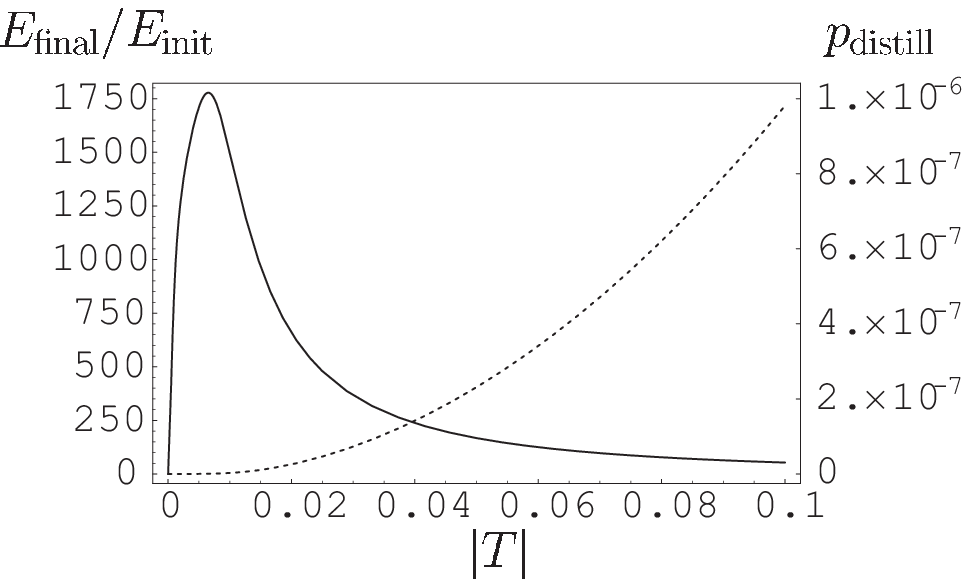}}
\caption{\label{fig:distillplot} This figure illustrates a full
distillation procedure. Beginning with a supply  of two-mode
squeezed vacua, with $q=0.01$, the protocol outlined in
Sec.~\ref{Procrustean} is then applied, which maps this state onto
a non-Gaussian state of higher entanglement, followed by three
iterations of the protocol described in Sec.~\ref{sec:protocol}. The properties of state
produced  depend on the transmittivity $T$ of the beam splitter
employed in the first step. Here, the factor by which the
entanglement of the final achieved state $E_{\text{final}}$ is
greater than the entanglement of the initial supply
$E_{\text{int}}$, (where the entanglement is calculated as the Von
Neumann entropy of the reduced density matrix of a single mode) is
plotted as a solid line, and the overall success probability of
the entire process, when this initial step is followed by three
iterations of the protocol to generate Gaussian states, is plotted as a dashed
line. }
\end{figure}

  An example of the results of  such a distillation
protocol, where the initial step is followed by three iterations
of the protocol from Sec.~\ref{sec:protocol}, is  illustrated in
Fig.~\ref{fig:distillplot}.  The overall probability is far lower
than for three steps of the protocol from Sec.~\ref{sec:protocol} alone (cf.
Fig.~\ref{fig:prob3}), due to the low success probability of the
initial step. This is largely due to the low probability of
measuring the presence of photons on the side  where no beam splitter is employed, i.e. Alice's
side. Since the effect of this measurement is to prepare a single
photon on Bob's, this low probability step  could be avoided
if a single photon source were available.

In light of the fact that distillation
with Gaussian operations alone was shown to be impossible \cite{Dist,Dist2}, it is then significant that
this scheme does, in fact, realise  pure-state distillation into \textit{approximate} Gaussian
states via suitable non-Gaussian operations, here photon detection.

  This simple protocol is not suitable when the initial supply consists of two-mode squeezed states with a high $q$, and another method of generating non-Gaussian states of higher entanglement must be used.  
A more detailed analysis of optimal preparation protocols that only
include passive optical elements and photon detectors will be
investigated elsewhere.
Here, we concentrate on the proof-of-principle
that Gaussian states can indeed be distilled to approximately Gaussian states.

%%%%%%%%%%%%%%%%%%%%%%%%%%%%%%%%%%%%%%%%%%%%%%%%%%%%%%%%%%%%%%%%%%%%%%
\section{Discussion and Conclusions}

We have shown that, using passive optical elements and photon
detectors that do not distinguish different photon numbers, one can
distill pure Gaussian states to arbitrarily high precision, in spite of
the impossibility of distilling Gaussian states with Gaussian
operations \cite{Dist,Dist2}. It should be noted that in our discussion we
have assumed the photon detectors to have unit efficiency, in order
to show that how one can in principle generate Gaussian states
 from a non-Gaussian supply.
Needless to say,
in any experimental realisation, one would have to deal with
detector efficiencies significantly
less than one. Such detectors can, e.g., be
modeled by employing perfect detectors, together with an appropriate
beam splitter with an empty input port \cite{Nemoto}.
If the detector efficiency is still close to one,
one would expect --  after a small number of iterations of the procedure --
the resulting states to be still close to those
presented in this idealised protocol. The convergence properties will in
general be different from the ideal situation. Dark counts of the detector,
in turn, do not affect the performance of the protocol, except that the success
probability is decreased. These matters will be discussed in more detail elsewhere.

In several practical applications of the procedure,
one can actually assume the initial state to be known. This is the
case, for example, if one uses the above protocol
in order to purify a state in a quantum privacy amplification
procedure \cite{HansHans}. Then, one may use homodyne detection
together with passive optical elements in order to implement a
positive-operator-valued measure (POVM)
$\{|\alpha\rangle\langle\alpha|:\,\alpha\in{\mathbbm{C}}\}$, where
$|\alpha\rangle$ denotes the state vector of a coherent state,
instead of photon detection \cite{Dist2,Leonhardt}. This would
render a displacement in phase space necessary in the last step,
depending on the measurement outcomes in each step. Such a
modification, however, would transform the originally
probabilistic protocol into a deterministic one. Also, the
detector efficiencies can be assumed to be significantly larger.
Even the displacement could be accounted for in the classical
analysis of the measured data in the final stage of a protocol
that makes use of the prepared entangled Gaussian state, e.g., a
quantum cryptography protocol.

In this article we have restricted our analysis to pure states. 
In practical implementations it would clearly also be useful 
to be able to distill highly entangled Gaussian staes from a 
mixed initial supply. However, the full treatment of these protocols 
for general mixed states is  lengthy and will be presented elsewhere.
To summarize, we have identified a procedure, that asymptotically produces Gaussian states
from a supply of non-Gaussian, finite-dimensional states   by
means of Gaussian operations  . In fact, the limiting Gaussian
state for a pure given input can be found analytically. We have
seen that even after a very small number of iteration steps the
degree of overlap between the resulting state and the theoretical
limit state is close to unity. Moreover, the probability of
obtaining this approximate state is of the order of 0.1. In that
respect the whole protocol is experimentally feasible with present
technology. This result should contribute to the search for
strategies to distribute continuous-variable entanglement over
large distances.

\acknowledgments

We would like to thank I.~Walmsley and J.I.~Cirac for fruitful discussions, and
K.~Audenaert, A.~Lund and W.J.~Munro   for very helpful remarks on the manuscript. This
work was partially funded by the Alexander~von~Humboldt
foundation, Hewlett-Packard Ltd., the EPSRC and the European
Commission (EQUIP).

%%%%%%%%%%%%%%%%%%%%%%%%%%%%%%%%%%%%%%%%%%%%%%%%%%%%%%%%%%%%%%%%%%%%%%

%\end{multicols}


\begin{thebibliography}{99}
\bibitem{Gauss}
Gaussian states of systems with canonical degrees of freedom are those
states with a Gaussian Wigner function (or equivalently, a Gaussian
characteristic function) \cite{vogelwelsch}.
\bibitem{Gen}
Note that this convergence is pointwise convergence for the
coefficients specifying the state vector.
\end{thebibliography}
\end{document}